\documentclass[preprint2]{aastex}
\usepackage{graphicx}
\usepackage{epstopdf}

\shorttitle{Monitoring of GS 1826-238}
\shortauthors{Rodi et al.}

\begin{document}


\title{Hard X-ray Tail Discovered in the Clocked Burster GS \(1826-238\)}


\author{J. Rodi\altaffilmark{1,2}, E. Jourdain\altaffilmark{1,2}, and J. P. Roques\altaffilmark{1,2}}
\affil{\textsuperscript{1}Universit\'e de Toulouse; UPS-OMP; IRAP;  Toulouse, France\\
\textsuperscript{2}CNRS; IRAP; 9 Av. Colonel Roche, BP 44346, F-31028 Toulouse cedex 4, France\\}

 
\begin{abstract}
The LMXB NS GS \(1826-238\) was discovered by \textit{Ginga} in 1988 September.  Due to the presence of quasi-periodicity in the type I X-ray burst rate, the source has been a frequent target of X-ray observations for almost 30 years.  Though the bursts were too soft to be detected by \textit{INTEGRAL}/SPI, the persistent emission from GS \(1826-238\) was detected over 150 keV during the \( \sim 10 \) years of observations.  Spectral analysis found a significant high-energy excess above a Comptonization model that is well fit by a power law, indicating an additional spectral component.  Most previously reported spectra with hard tails in LMXB NS have had an electron temperature of a few keV and a hard tail dominating above \( \sim 50 \) keV with an index of \(  \Gamma \sim 2-3 \).  GS \(1826-238\) was found to have a markedly different spectrum with \( kT_e \sim 20 \) keV and a hard tail dominating above \( \sim 150 \) keV with an index of \( \Gamma \sim 1.8\), more similar to BHXRB.  We report on our search for long-term spectral variability over the \(25-370\) keV energy range and on a comparison of the GS \(1826-238\) average spectrum to the spectra of other LMXB NS with hard tails.
\end{abstract}




\section{Introduction}
GS \(1826-23\) was discovered by \textit{Ginga} on 1988 September 8 \citep{makino1988} and was initially considered to be a black hole candidate because its hard spectrum and flickering were similar to Cyg X-1 \citep{tanaka1989}.  Not until \citet{ubertini1997} were X-ray bursts reported from the source to reveal the compact object as a neutron star (NS).  \citet{homer1998} confirmed this result by detecting optical bursts.  They also measured a 2-day orbital period for GS \(1826-238\).  With \( \sim 2\) Ms of \textit{BeppoSAX} exposure time, \citet{ubertini1999} observed 70 X-ray bursts and found a quasi-periodicity in the burst recurrence time of 5.76 h, which suggests a stable accretion rate.  

Observations by the ROSAT-PSPC identified the optical counterpart, determining that the system is a low mass X-ray binary (LMXB) \citep{motch1994,barret1995}.  From \citet{barret1995} the estimated distance to the source is \(4-10\) kpc, while \citet{int1999} and \citet{kong2000} place an upper limit of 8 kpc on the distance.  

LMXB NS systems can be broadly classified as either Z-sources or atoll sources.  Z-sources trace out a 'Z' shape in a color-color diagram while atoll sources trace out a 'banana' shape \citep{church2014}.  \citet{muno2002} tentatively classified GS \(1826-238\) as an atoll source though the source had shown very little variability in its color-color diagram over the 5 years of observations they studied.

Previous X-ray observations of GS \(1826-238\) have found similar behavior, with the source consistently in a hard spectral state with flux out to 100 keV \citep{strickman1996,int1999,barret2000,thompson2005,cocchi2010,cocchi2011}.  The high-energy sensitivity of SPI above 100 keV and the numerous observations of the Galactic Center region by \textit{INTEGRAL} allow for a study of the high-energy spectrum of this source above 100 keV.

In this work, we looked at the hard X-ray/soft gamma-ray spectrum of GS \(1826-238\) {out to \(370\) keV having analyzed \( \sim 11 \) Ms of SPI data over the span of \( \sim 3900 \) days.  We searched for temporal and spectral variability from the source on timescales longer than 3 days.  Finally, we compared the average GS \(1826-238\) spectrum with the spectra of other LMXB NS that have been reported to have hard tails, specifically the atoll source 4U 1728-34.

\begin{table*}
\begin{center}
\caption{\textit{INTEGRAL}/SPI observations of GS \(1826-238\)}
%\caption{More terribly relevant tabular information.\label{tbl-2}}
\begin{tabular}{cccc}
\tableline\tableline \\
Period & \textit{INTEGRAL} Revolutions & Time                 & Exposure Time\\
       &                               & (MJD)                &    (ks)      \\  
\tableline
1      &  \(0053-0065\)                & \(52719-52757\)      &    238       \\
2      &  \(0105-0122\)                & \(52874-52928\)      &    1249      \\
3      &  \(0165-0185\)                & \(53054-53116\)      &    501       \\
4      &  \(0225-0246\)                & \(53233-53298\)      &    745       \\
5      &  \(0286-0310\)                & \(53416-53490\)      &    509       \\
6      &  \(0348-0371\)                & \(53601-53672\)      &    666       \\
7      &  \(0408-0431\)                & \(53780-53851\)      &    805       \\
8      &  \(0472-0495\)                & \(53972-54042\)      &    604       \\
9      &  \(0534-0552\)                & \(54157-54213\)      &    547       \\
10     &  \(0594-0606\)                & \(54337-54375\)      &    565       \\
11     &  \(0661-0669\)                & \(54537-54563\)      &    197       \\
12     &  \(0724-0736\)                & \(54725-54764\)      &    272       \\
13     &  \(0776-0795\)                & \(54881-54940\)      &    316       \\
14     &  \(0838-0846\)                & \(55066-55090\)      &    138       \\
15     &  \(0908-0918\)                & \(55276-55308\)      &    115       \\
16     &  \(1025-1036\)                & \(55626-55661\)      &    276       \\
17     &  \(1080-1094\)                & \(55790-55835\)      &    300       \\
18     &  \(1145-1157\)                & \(55985-56023\)      &    471       \\
19     &  \(1203-1226\)                & \(56158-56228\)      &    942       \\
20     &  \(1261-1282\)                & \(56332-56396\)      &    513       \\
21     &  \(1326-1348\)                & \(56526-56594\)      &    786       \\
\tableline
Total: &                              &                      &    10 755    \\
\tableline\tableline
\end{tabular}
%% Any table notes must follow the \end{tabular} command.
%\tablenotetext{a}{Sample footnote for table~\ref{tbl-2} that was
%generated with the \LaTeX\ table environment}
%\tablenotetext{b}{Yet another sample footnote for table~\ref{tbl-2}}
%\tablenotetext{c}{Another sample footnote for table~\ref{tbl-2}}
%\tablecomments{We can also attach a long-ish paragraph of explanatory
%material to a table.}
\label{tab:obs}
\end{center}
\end{table*}

\section{Instrument and Observations}
On 2002 October 17, the \textit{International Gamma-ray Astrophysics Laboratory} (\textit{INTEGRAL}) was launched from Baikonur, Kazachstan.  The satellite has an eccentric \( \sim 3\)-day orbital period and an inclination of \(51.6^{\circ}\) \citep{jensen2003}.  The spectrometer \textit{INTEGRAL}/SPI spans the \(20\) keV \(-\) 8 MeV energy range with an energy resolution of \( 2-8 \) keV \citep{roques2003}.  \textit{INTEGRAL} has observed the Galactic Center region at approximately 6 month intervals and thus has created a long observational baseline for studying sources in this region.

\begin{figure*}[t]
  \centering
  \includegraphics[scale=0.7, angle=180,trim = 20mm 20mm 0mm 10mm, clip]{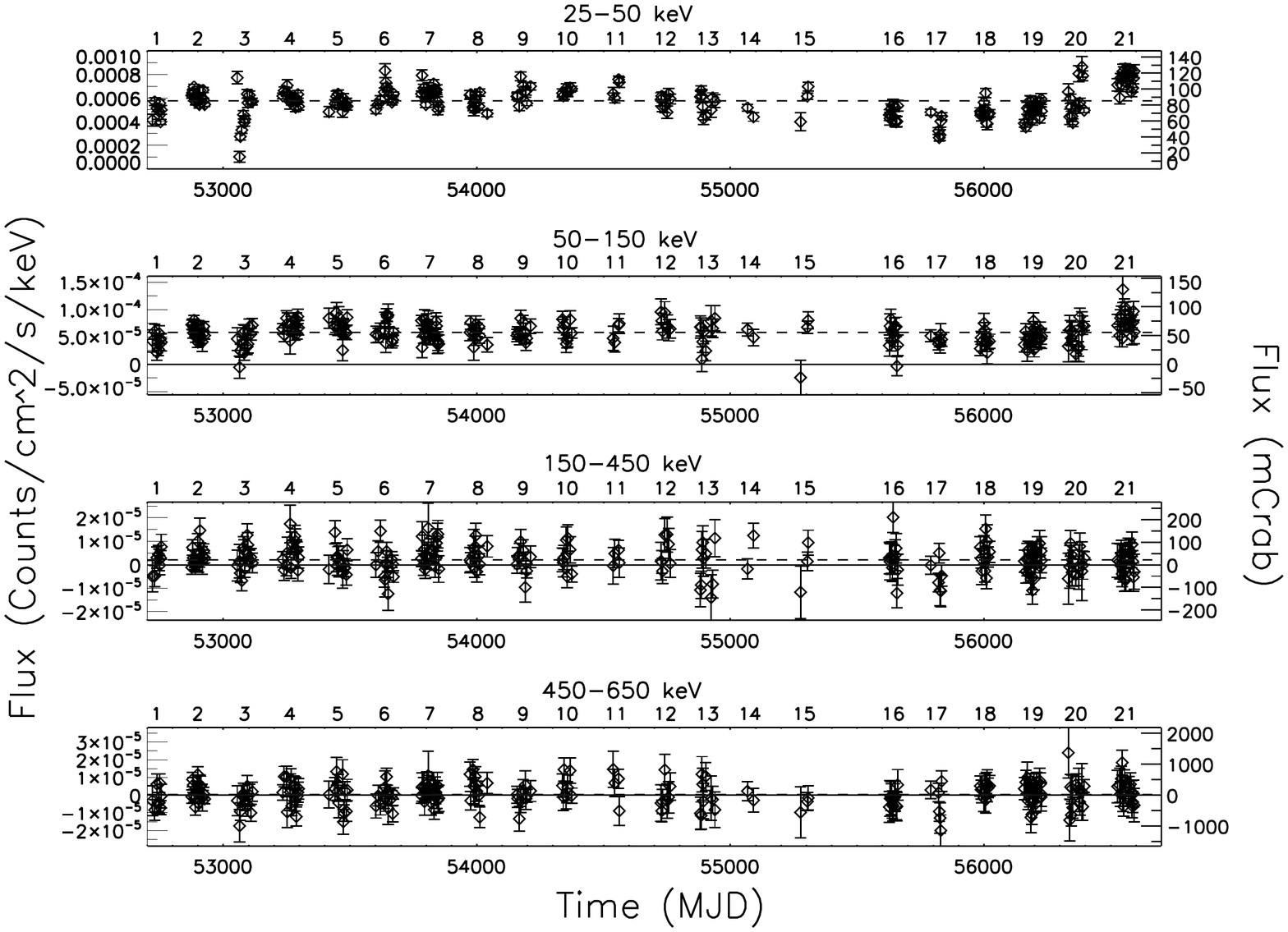}
  \caption{SPI light curve for GS \(1826-238\) spanning MJD \(52700-56700\) in 4 broad energy bands from \(25-650\) keV.  The dashed line in each panel corresponds to the average flux, and the solid line corresponds to a flux of 0 mCrab.  The Period number has been plotted above the corresponding data for each panel.} \label{fig:lc}
\end{figure*}

We have utilized this long baseline by analyzing 223 \textit{INTEGRAL} revolutions spanning MJD \(52719 - 56594\) (2003 March 21 \(-\) 2013 October 29) using the SPI Data Analysis Interface (SPIDAI\footnote{Publicly available interface developed at IRAP to analyze SPI data. Available at http://sigma-2.cesr.fr/integral/spidai. See description in \citet{burke2014}}) using only the revolutions where GS \(1826-238\) was within \( 12^{\circ} \) of the SPI pointing direction for at least 10 science windows (\( \sim 2000 \) s per science window).  This resulted in a total exposure time of over 10 Ms.  Table~\ref{tab:obs} lists the range of \textit{INTEGRAL} revolutions during each observation period, the beginning and ending MJD of the period and the exposure time during the period.  

Because GS \(1826-238\) is in the Galactic Center region, there will be some contribution to the detector count rate due to diffuse positronium emission above \( \sim 300 \) keV, which could potentially result in an artificial high-energy spectral tail for the source.  To mitigate this effect, the expected incident count-rate due to diffuse positronium was modeled based on parameters reported by \citet{bouchet2008} and then subtracted from the raw data before analyzing the data with SPIDAI.

In the spectral analysis, the SPI data were grouped into 50 energy bins spanning \(22-650\) keV.  The first two energy channels were ignored in spectral fitting because of uncertainties in the energy response \citep{jourdain2009} thus reducing the energy range to \(25-650\) keV.  No significant long-term emission was detected above 370 keV thus the spectral analysis has been limited to \(25-370\) keV.

\section{Results}
\subsection{Temporal Variability}

GS \(1826-238\) exhibits frequent type I X-ray bursts with  a quasi-periodicity of 5.76 h \citep{ubertini1999}.  The duration of these bursts are typically \( \sim 100 \) s in the \(2-10\) keV energy range \citep{ji2014} and are soft enough to have a negligible effect \( > 25 \) keV.  GS \(1828-238\) also undergoes longer duration variability, which can be seen in Figure~\ref{fig:lc}.  The SPI light curve is shown for four broad energy bands (\(25-50\) keV, \(50-150\) keV, \(150-450\) keV, and \(450-650\) keV) covering MJD \(52700-56700\) with each point corresponding to the average flux during that \textit{INTEGRAL} revolution.  In each panel the solid line denotes a flux of 0 mCrab (except for the top panel), and the dashed line denotes the long-term average flux in that energy band.  For each panel, the Period numbers have been plotted above the corresponding data.  The total \(25-50\) keV significance is \(265.1 \sigma \), the \(50-150\) keV significance is \(65.2 \sigma\), the \(150-450\) keV significance is \(7.0 \sigma\), and the \(450-650\) keV significance is \(1.2 \sigma\).  GS \(1826-238\) shows relatively little variability within a period, apart from Period 3.  During this period, the \(25-50\) keV flux decreased from \( \sim 120\) mCrab to below 20 mCrab in \( \sim 5 \) days before increasing to \( \sim 90\) mCrab in approximately 40 days.

GS \(1826-238\) has undergone two large dips after this work (during \textit{INTEGRAL} revolutions not currently publicly available).  A dip began on MJD 56816 (2014 June 8) when \textit{MAXI} detected a decrease in the hardness ratio from \( \sim 0.4 \) to 0.1 along with the \(2-10\) keV flux increasing from roughly 50 mCrab to 140 mCrab on MJD 56822 (2014 June 14) \citep{nakahira2014}.  During this time the \(15-50\) keV \textit{Swift}/BAT flux decreased, as can be seen in the publicly available light curve\footnote{http://swift.gsfc.nasa.gov/results/transients/Ginga1826-238/}.  Based on the BAT data, the flux decreased from \( \sim 110 \) mCrab to \( \sim 5\) mCrab over about 5 days, remaining in a low flux state until \( \sim \) MJD 56850.  After which, the flux began to increase over the span of about 40 days, and finally recovered to \( \sim 100 \) mCrab around MJD 56890 (2014 August 21).  

In the BAT light curve, the next dip can be seen to start about MJD 57150 (2015 May 8).  This time the flux decreased over the span of \( \sim 25\) days from \( \sim 100 \) mCrab to fluxes consistent with 0 mCrab.  The source spent only a few days in this low flux state before recovering to the initial flux level over the course of about 30 days until \( \sim \) MJD 57210 (2015 July 7).
  
The first two dips show similar timescales of a few days for the flux decreases.  The third dip begins with a slow decline over \( \sim 20\) days, before starting a fast decline lasting a few days.  All of the dips reached a minimum flux of near or consistent with 0 mCrab.  Also, the timescales for the flux increases are roughly \(30-40\) days long.  The durations of the dips also show a difference, with the second burst having lasted a few 10's of days near its minimum flux level while the first and last dips spent only a few days at their minimum flux levels.

\begin{figure*}[t!]
  \centering
  \includegraphics[scale=0.6, angle=180,trim = 30mm 35mm 20mm 20mm, clip]{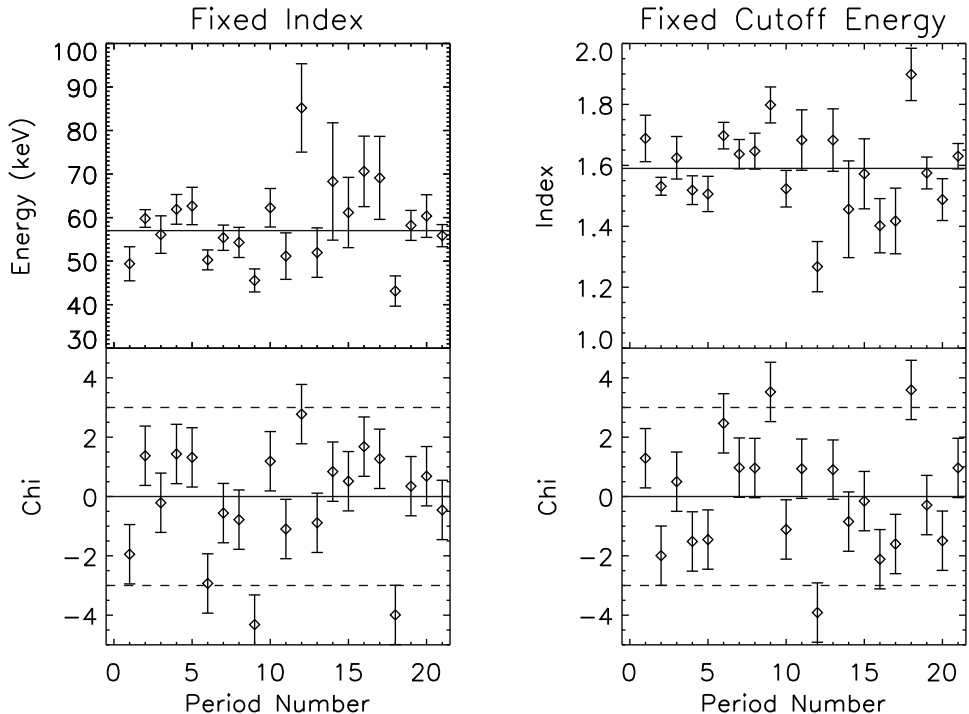}
  \caption{\textit{Left Column:} (\textit{Top}) Plot of cutoff energy for each period when the photon index is held fixed at \( \Gamma = 1.59\).  Solid line denotes best-fit cutoff energy from average spectrum (\(E_{cut} = 57\) keV).  (\textit{Bottom}) Plot of significance of period cutoff energy from average spectrum.  Solid line denotes deviation of \(0 \sigma\), and dashed lines denote deviations of \(+ 3 \sigma \textrm{ and } - 3 \sigma\).  \textit{Right Column:} (\textit{Top}) Plot of photon index for each period when the cutoff energy is held fixed at \(E_{cut} = 57\)  keV.  Solid line denotes the best-fit photon index from average spectrum (\( \Gamma = 1.59\)).  (\textit{Bottom}) Plot of significance of period photon index from average spectrum.  Solid line denotes deviation of \(0 \sigma\), and dashed lines denote deviations of \(+ 3 \sigma \textrm{ and } - 3 \sigma\).} 
\label{fig:specvar}
\end{figure*}

\subsection{Spectral Variability}

Due to a low signal-to-noise ratio during a single revolution, a search for spectral variability on a period timescale was performed.  To do so, an average spectrum was generated using all periods except Period 3, which shows the large flux variability.  The average spectrum was poorly fit by a power law model with a best-fit photon index of \( \Gamma = 2.52 \pm 0.01 \) with \( \chi^2 /  \nu = 11.80\) (\( \nu = 36\)), but was significantly better fit by a cutoff power law with \( \Gamma = 1.59 \pm 0.05 \) and a cutoff energy of \( E_{cut} = 57 \pm 3 \) keV with \( \chi^2 /  \nu = 1.56\) (\( \nu = 35 \)) with residuals suggesting a high-energy excess above \( \sim 150 \) keV.  To compare the period spectra with the average spectrum, each period spectrum was fit once with the photon index fixed to the average value (\(\Gamma = 1.59\)), and the cutoff energy was allowed to vary.  Then the data were fit again with the cutoff energy fixed (\( E_{cut} = 57\) keV), and the photon index was allowed to vary.  

\begin{figure}[h!]
  \centering
  \includegraphics[scale=0.8, angle=180,trim = 130mm 51mm 27.5mm 15mm, clip]{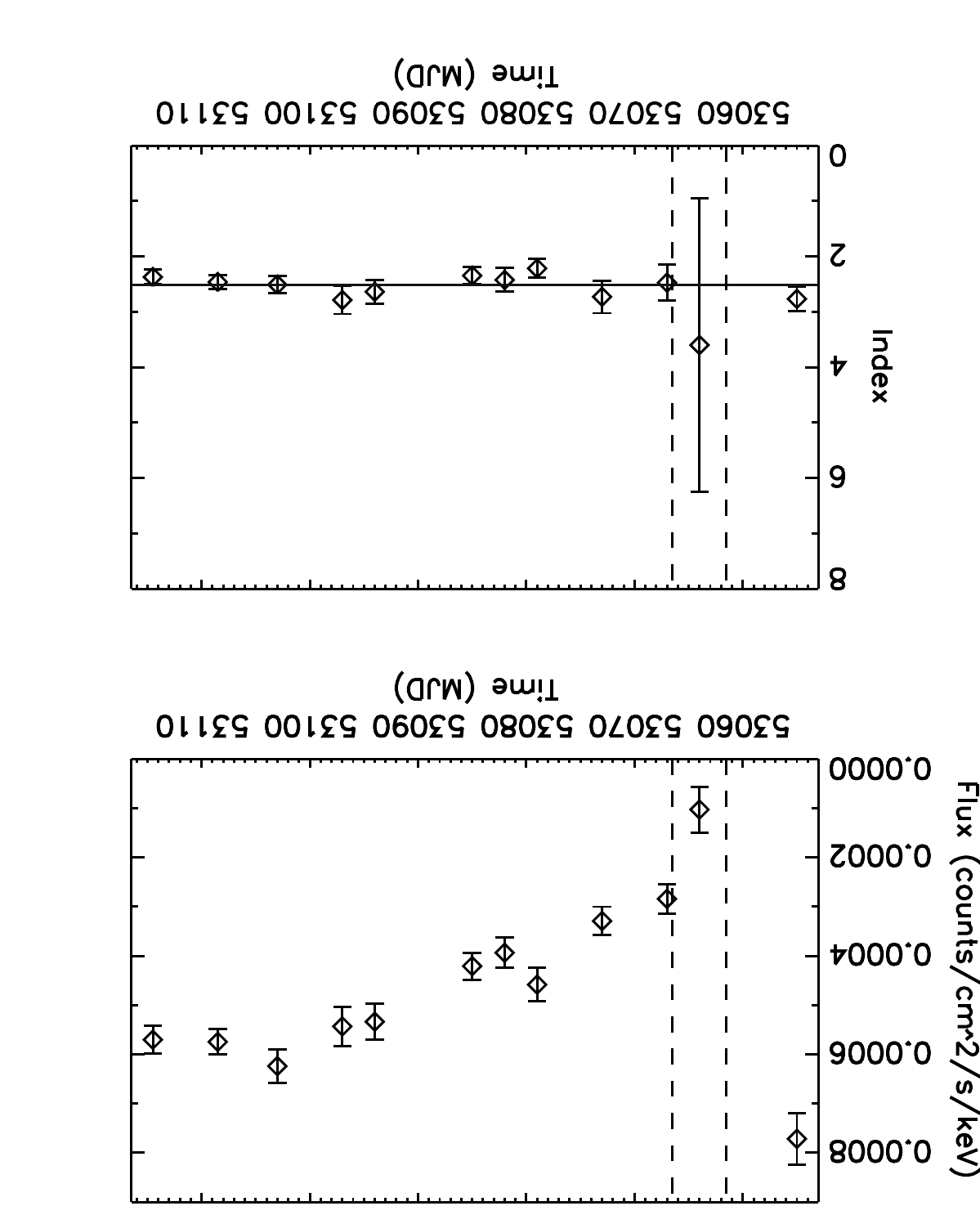}
  \caption{\textit{Top}: Plot of \(25-50\) keV SPI light curve for Period 3 revolutions.  \textit{Bottom}:  Plot of photon index for Period 3 revolutions.  Solid line denotes the photon index of the average spectrum with \( \Gamma = 2.52\).  Dashed vertical lines in both panels denote start and stop times of potential soft state based on ASM data.}
\label{fig:per3}
\end{figure}

\label{sec:hepo}
  
\begin{figure}[h!]
  \centering
  \includegraphics[scale=1.0, angle=0,trim = 0mm 130mm 125mm 29mm, clip]{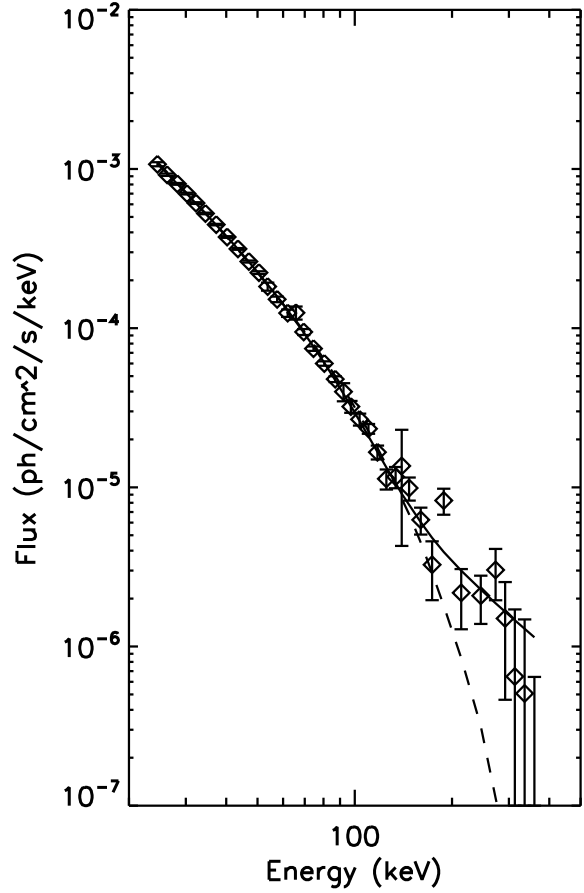}
  \caption{Average \(25-370\) keV spectrum.  \texttt{CompTT} model shown with dashed line.  \texttt{CompTT \(+\) powerlaw} model shown with solid line. } 
\label{fig:spec}
\end{figure}

In the left column of Figure~\ref{fig:specvar}, the top panel shows the cutoff energy for each period when the photon index is fixed with the solid line marking the cutoff energy from the average spectrum.  The bottom panel shows how significantly the cutoff energies deviate from the long-term average.  The solid line denotes a value of \(0 \sigma \) away from the average value while the dashed lines denote \( + 3 \sigma \textrm{ and } -3 \sigma\) away from the average value.  The top panel in the right column shows the photon index when the cutoff energy is fixed with the bottom panel again showing the significance of the deviation from the long-term average.

When the photon index is fixed, only two periods have a cutoff energy \( > |3 \sigma| \) away from the average.  The two outlying periods are 9 and 18, and both are \( \sim 4 \sigma \) below the average.  Similarly, when the cutoff energy is fixed, these two periods again show a large deviation from the average (\( \sim 3.5 \sigma \) above the average).  Also, a third period is \( \sim 4 \sigma \) below the average, Period 12.  When the photon index is fixed, Period 12 has a deviation of a little under \( + 3 \sigma \).  As Periods 9, 12, and 18 show deviations \( > |3 \sigma| \) away from the average with either the photon index fixed or the cutoff energy fixed, these periods have been removed from the long-term analysis and analyzed separately.  (See below.)  

\subsubsection{Peculiar Periods}
Based on the top plot in the right column of Figure~\ref{fig:specvar}, Periods 9 and 18  have photon indexes that are relatively soft compared to the other periods while the photon index for Period 12 is harder than all of the other periods.  When the cutoff energy is fixed to 57 keV, the photon indexes for Periods 9, 12, and 18 are \( 1.80 \pm 0.06 \) (\( \chi^2 / \nu = 1.17\)), \(1.27 \pm 0.08\) (\(1.07\)), and \(1.90 \pm 0.09\) (\(1.02\)), respectively.  The indexes for Periods 9 and 18 are marginally consistent and were combined, resulting in a photon index of \( \Gamma = 1.83 \pm 0.05\) (\( \chi^2 / \nu = 0.82\)).  Periods 9 and 18  correspond to a spectral softening compared to the average spectrum while Period 12 corresponds to a spectral hardening.

Interestingly, Period 3, which shows the large flux dip in the \(25-50\) keV light curve (Figure~\ref{fig:lc}), has fit values consistent with the average spectrum in both cases.  Such a dip is suggestive of a hard-to-soft state transition.  Analysis of the \(2-12\) keV data from the \textit{RXTE}/All-Sky Monitor (ASM) shows a possible hard-to-soft transition on roughly MJD 53061.5 with a possible soft-to-hard transition on about MJD 53066.5.  During this 6 day time period, the ASM data are \( \sim 5.5 \sigma\) above the long-term average flux, but the source variability seen outside this MJD range and the relatively low significance make a confident detection of a state change difficult.

The interpretation of a short lived soft state is not inconsistent with the SPI data.  In this scenario, the hard-to-soft transition occurs \(\sim 2\) days before the beginning of revolution 0168, and the soft-to-hard transition occurs about 0.5 days after the start of revolution 0169.  The start and stop times are marked in both panels of Figure~\ref{fig:per3} with dashed lines.  The top panel of Figure~\ref{fig:per3} shows the \(25-50\) keV light curve for Period 3 while the bottom panel shows the best-fit photon index for each revolution.  In the bottom panel, the solid line denotes the index of the long-term average spectrum \(\Gamma = 2.52\).

When fit to a power law model, all of the revolutions have an index of \( \Gamma \sim 2.5\) except for revolution 0168 (the lowest flux revolution), which has an index of \( \Gamma \sim 4\) though with large errors due to its low flux.  Because revolution \(0168\) shows signs of a soft state, these data were removed from further analysis.

\subsubsection{Average Spectrum}
The updated spectrum with Periods 9, 12, and 18 and revolution \(0168\) from Period 3 removed had an exposure time of 9.5 Ms.  As before, a cutoff power law model resulted in a large \( \chi^2 /  \nu\) (1.41) with residuals above \( \sim 150\) keV, suggesting a high-energy component.  Thus the data were fit with a cutoff powerlaw \(+\) powerlaw model.  This model greatly decreased the \( \chi^2 / \nu\) to 0.81 with parameters \( \Gamma_1 = 1.22 \pm 0.30 \textrm{, } E_{cut} = 35 \pm 10 \textrm{ keV, and } \Gamma_2 = 1.80 \pm 0.98 \).  An F-test resulted in a probability of chance improvement of \(3.96 \times 10^{-5}\), which corresponds to \(4.1 \sigma\).  

For a physical interpretation, the data were fit to a \texttt{CompTT} Comptonization model \citep{titarchuk1994} assuming spherical accretion geometry resulting in model parameters with an electron temperature of \( kT_e = 26 \pm 1 \) keV and an optical depth of \( \tau = 3.21 \pm 0.16 \) with \( \chi^2 /  \nu = 2.12\) (\(\nu = 35\)) with the high-energy residuals still present.  Adding a high-energy power law component yields fit parameters of \( kT_e = 18 \pm 2 \textrm { keV, } \tau = 4.39 \pm 0.42 \textrm{, and } \Gamma = 1.77 \pm 0.61 \) with \( \chi^2 /  \nu = 0.87\).  From an F-test, the probability of chance improvement was \( 1.63 \times 10^{-7} \), which corresponds to \( 5.2 \sigma \).  Thus the addition of a high-energy component is statistically significant.  

The data were also fit to \texttt{CompTT} model convolved with a reflection component \citep{magdziarz1995} with the reflection fraction fixed to 1.  The fit with a \texttt{reflect(compTT)} model resulted in fit parameters with \( kT_e = 34 \pm 2 \) keV and \( \tau = 2.61 \pm 0.21 \) (\(\chi^2 / \nu = 1.39\), \(\nu = 35\)).  The electron temperature is significantly higher than the \texttt{CompTT} alone, but, like the \texttt{CompTT} model alone, this model also fails to adequately fit the high-energy tail.

Figure~\ref{fig:spec} shows the average spectrum plotted with the \texttt{CompTT} model shown with a dashed line and the \texttt{CompTT \(+\) powerlaw} model shown with a solid line.  From the plot, it is clear that the \texttt{CompTT} model fails to describe the data above \( \sim 150 \) keV.  As the positronium emission has been removed from the data, the hard tail is likely the result of emission from GS \(1826-238\).

\section{Discussion}
\subsection{Comparison with Previous Observations}
\subsubsection{Previous \textit{INTEGRAL} Observations}
The interesting bursting behavior of GS \(1826-238\) means that it has been observed in X-rays often since its discovery.  This allowed for several comparisons with the SPI results.  \citet{cocchi2010} studied GS \(1826-238\) in the \(3-200\) keV energy range using \textit{INTEGRAL}/JEM-X and ISGRI over three time periods from 2003 April to 2006 November.  These observations were during \textit{INTEGRAL} revolutions \(0061-0064\), \(0119-0122\), and \(0495\) with data during bursts removed. 

\citet{cocchi2010} found the \(2-200\) keV flux to be roughly constant when comparing revolutions \(0061-0064\) (MJD \(52743-52754\)) and \(0119-0122\) (MJD \(52916-52928\)).  During revolution \(0495\) (MJD 54041), the flux was observed to be \( \sim 30\% \) lower.  The science windows within \( 5^{\circ} \)  of the SPI pointing direction were used to match the field of view of JEM-X and thus use the same observations as \citet{cocchi2010}.  Analysis of the \(25-200\) keV SPI data found results in agreement with those from \citet{cocchi2010}.

The SPI data during these intervals were fit to a cutoff power law model over the \(25-200\) keV energy range to best match the \citet{cocchi2010} energy range.  Figure~\ref{fig:contour} shows the contour plots for each interval with the best fit parameters by a black 'X' for SPI and by a black square for \citet{cocchi2010}.  The contours mark the \(68.3\%\) (black), \(90\%\) (red), and \(99\%\) (green) confidence regions.

The best fit parameters to the SPI data for revolutions \(0061-0064\), \(0119-0122\), and \(0495\) are: \( \Gamma = 1.05 \textrm{ and } E_{cut} = 32 \) keV (\( \chi^2 /  \nu = 1.30\), \( \nu = 28\)), \( \Gamma = 1.61 \textrm{ and } E_{cut} = 55 \) keV (\( \chi^2 /  \nu = 0.58\)), and \( \Gamma = 1.11 \textrm{ and } E_{cut} = 27 \) keV (\( \chi^2 /  \nu = 0.88\)), respectively.  As shown in the left panel, the results from \citet{cocchi2010} are within the \(90 \%\) confidence region.  For the other two panels, the \citet{cocchi2010} results are within the \( 68.3 \%\) confidence regions, though for revolution \(0495\) that region is large due to a low signal-to-noise ratio.  Thus these SPI results are consistent with those from \citet{cocchi2010}.

\begin{figure*}
  \centering
  \includegraphics[scale=0.6, angle=180,trim = 0mm 30mm 20mm 90mm, clip]{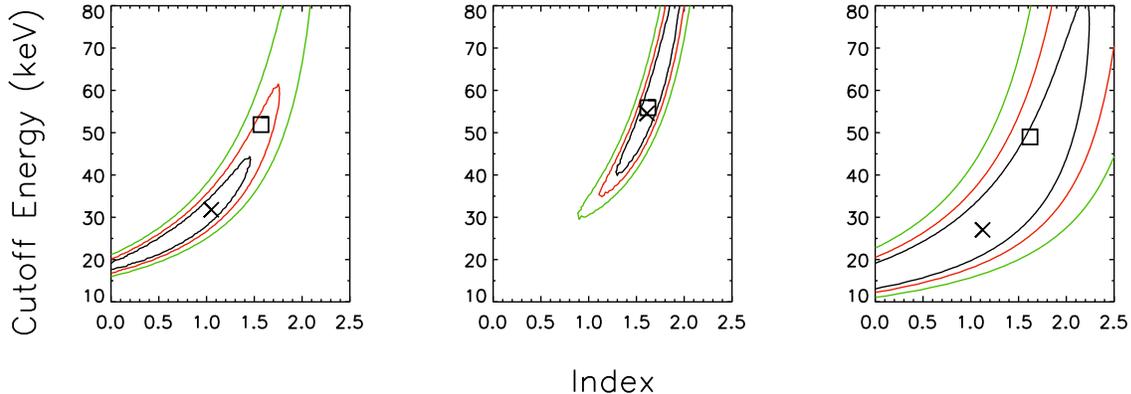}
  \caption{Confidence regions in the \(E_{cut} - \Gamma\) parameter space for \textit{INTEGRAL} revolutions \(0061-0064\) (\textit{left}), \(0119-0122\) (\textit{center}), and \(0495\) (\textit{right}) with contours for the \(68.3 \%\) (black), \(90 \%\) (red), and \(99 \%\) (green) confidence regions.  The best fit parameters from SPI are plotted with a black 'X' and the best fit parameters from \citet{cocchi2010} are plotted with a black square.}
\label{fig:contour}
\end{figure*}
 
\subsubsection{\textit{BeppoSAX} and OSSE Observations}
\citet{int1999} analyzed the \( 0.1-200 \) keV \textit{BeppoSAX} data covering 1997 April \(6.7-7.2\).  The persistent emission was fit to a blackbody + cutoff powerlaw model.  The photon index found was \( \Gamma = 1.38 \pm 0.03 \), and the cutoff energy found was \( E_{cut} = 51.69 \pm 0.03\) keV.  When the long-term average SPI spectrum was fit to the \(25-200 \) keV energy range, the cutoff power law model showed a steeper spectral index of \( \Gamma = 1.56\) with a  \( E_{cut} = 56 \pm 3 \) keV, and the \( \chi^2/ \nu = 1.22 \) (\( \nu =28\)).  Figure~\ref{fig:int} shows the confidence regions for the SPI fit using the same contour levels as in Figure~\ref{fig:contour}.  The fit parameters from \citet{int1999} are shown by a black square and are well outside the \(99 \%\) confidence region, indicating spectral variability between the SPI observations and the \textit{BeppoSAX} observations.

Additionally, \citet{int1999} fit the \( 60-150\) keV data to a powerlaw to compare with OSSE observations \citep{strickman1996} in the \(60-300\) keV range.  The \textit{BeppoSAX} data were best fit by a photon index of \( \Gamma = 3.3 \pm 0.4 \) and the OSSE data by an index of \( \Gamma = 3.1 \pm 0.5 \).  The \( 60-370\) keV SPI data are well fit by a photon index of \( \Gamma = 3.11 \pm 0.07\), in agreement with both results, indicating that the hard X-ray/soft gamma-ray spectrum has remained relatively constant for almost 20 years despite variability at lower energies.

\begin{figure}
  \centering
  \includegraphics[scale=0.6, angle=180,trim = 0mm 30mm 20mm 90mm, clip]{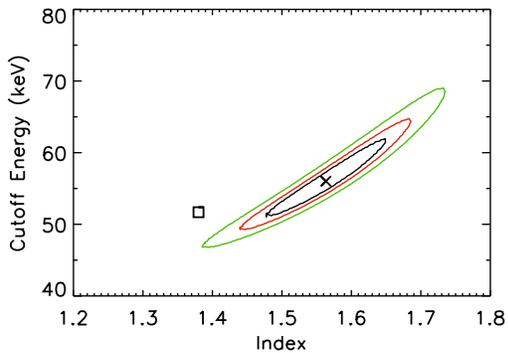}
  \caption{Contour plot showing the confidence regions for the SPI long-term average spectrum fit to a cutoff power law in the \(25-200\) keV energy range.  Contours mark the \(68.3 \%\) (black), \(90 \%\) (red), and \(99 \%\) (green) confidence regions.  The \textit{BeppoSAX} fit parameters from \citet{int1999} are marked by a black square.}
\label{fig:int}
\end{figure}

\subsection{Comparison with Other LMXB NS Hard Tails}

Since the discovery of a hard tail in GX 5-1 with \textit{Ginga} \citep{asai1994}, observations of LMXB NS have found that similar hard tails are not uncommon (e.g. Sco X-1 \citep{disalvo2006,dai2007,revnivtsev2014}, GX 17+2 \citep{disalvo2000}, 4U \(1636-53\) \citep{fiocchi2006}, and 4H \(1820-30\) \citep{tarana2007}) and have been seen in both Z-sources (Sco X-1 and GX 17+2) and atoll sources (4U 1636-53 and 4H 1820-30).  A majority of the spectra showing tails can be characterized by a soft state with a Comptonization component with a temperature of \(kT_e \sim 3\) keV and a power law excess above \( \sim 20-50 \) keV with a slope of \( \Gamma \sim 2-3\) and extends up to at least 100 keV.

As shown in Figure~\ref{fig:spec}, the spectrum of GS \(1826-238\) is considerably different with the source in a hard state with a Comptonization component extending out to \( \sim 150 \) keV  after which the hard tail becomes the dominant component with \( \Gamma = 1.77\).  Due to the large error, the photon index is consistent with the lower portion of the \( \Gamma \sim 2-3\) range.

Observations by \citet{tarana2011} found the atoll source 4U \(1728-34\) in the hard state with spectra similar to GS \(1826-238\).  When fitting the data to a \texttt{CompTT + powerlaw } model, the electron temperatures for these spectra were \( kT_e \sim 5.7 \textrm{, } 9.1 \textrm{, and } 10.1 \) keV with power law indexes of \( \Gamma \sim 2.5 \textrm{, } 1.9 \textrm{, and } 1.8\).  The errors on the tails with \( \Gamma < 2 \) were large enough to still be consistent with \( \Gamma \sim 2-3\). Thus the hard tails observed in both the soft and hard states are likely to be the same with the hard tail ``hidden'' by the Comptonization component when the electron temperature increases.

\citet{tarana2011} found that the hard states of 4U \(1728-34\) could be well fit using a hybrid thermal/non-thermal \texttt{CompPS} model \citep{poutanen1996}.  The fit parameters from the three observations were: (1) \( kT_e \sim 27 \) keV, \( \tau_y \sim 3\), and \( p \sim 3.2\), (2) \( kT_e \sim 24 \) keV, \( \tau_y \sim 2.8\), and \( p \sim 1.4\), (3) \( kT_e \sim 20 \) keV, \( \tau_y \sim 2.4\), and \( p \sim 0.6\).  The GS 1826-238 average spectrum was fitted to the same model, resulting in best fit parameters of \( kT_e = 31 \pm 2 \) keV, an optical depth of \( \tau_y = 2.56 \pm 0.23\), and an electron power-law index \(p = 0.65 \pm 0.27\) with \( \chi^2 / \nu = 0.82 \) (\( \nu = 34\)).  The GS \(1826-238\) fit has a slightly higher electron temperature with a slightly lower optical depth and a hard electron power-law index similar to spectrum (3) from 4U \(1728-34\).

\section{Conclusion}
In this work, we presented \( \sim 11 \) Ms of GS \(1826-238\) observations from \textit{INTEGRAL}/SPI that span \( \sim 10.5\) years from \(2003-2013\).  The SPI results show long-term flux variability with a dramatic decrease in the \(25-50\) keV flux around MJD 53000 with a possible short-lived transition to a soft state before recovering to near pre-dip levels.  Spectral analysis showed little variability outside of Periods 9, 12, and 18 and revolution 0168 in Period 3.  (See Section 3.2.1.)  This allowed us to look at 9.5 Ms of exposure time in the average spectrum which spanned the \(25-370\) keV energy range.

The spectrum was dominated by a Comptonization component up to \( \sim 150\) keV with a high-energy excess extending up to \( \sim 400\) keV.  Fitting the data to either a cutoff power law or a \texttt{CompTT} model required a power law tail to achieve an acceptable  \( \chi^2 / \nu \).  For the cutoff power law model the data were well described by \( \Gamma_1 = 1.22 \textrm{, } E_{cut} = 35 \textrm{ keV, and } \Gamma_2 = 1.80\).  In the case of the \texttt{CompTT} model, the best fit parameters were  \( kT_e = 18 \textrm{ keV, } \tau = 4.39 \textrm{, and } \Gamma = 1.77\).  The spectrum was also well fit by a hybrid thermal/non-thermal \texttt{CompPS} model with  \( kT_e = 31 \) keV, \(p = 0.65 \), and \( \tau_y = 2.56\).

Comparison with \textit{BeppoSAX} results from \citet{int1999} found a significantly harder photon index than the SPI index in the \(25-200 \) keV range, but found a similar cutoff energy for a cutoff power law model.  Also, the comparison of the spectra \( > 60 \) keV between OSSE, \textit{BeppoSAX}, and SPI found all in agreement with the data well fit by a power law of \( \Gamma \sim 3.1\).  These results indicate that the spectral shape remained roughly constant around 100 keV while varying at lower energies over the span of nearly 20 years.

A comparison of the GS \(1826-238\) spectrum to the spectra of other NS with hard tails found GS \(1826-238\) to have a significantly hotter electron temperature, resulting in the hard tail not becoming the significant component until \( \sim 150\) keV.  Similarities to 4U \(1728-34\) were found as both sources  show a high energy component extending out above at least \( \sim 100 \) keV.  Both sources could be well fit using the hybrid thermal/non-thermal Comptonization model, \texttt{CompPS}, which indicates the presence of a non-thermal process at work at high energies.

The presence of a non-thermal hard tail is often interpreted as being related to a jet \citep{markoff2005,migliari2010,tarana2011}.   Correlations between radio and X-ray fluxes have been reported for 4U \(1728-34\) \citep{migliari2003} and GX \(17+2\) \citep{migliari2007}, which supports that interpretation.  As of yet no detection of GS \(1826-238\) has been reported, but \citet{fender2000} estimate the radio emission in the hard state to be at least \(5-10\) fainter than the soft state.  With the exception of the possible short-lived soft state around MJD 53000, the only reported soft state observations of GS \(1826-238\) have been by MAXI in 2014 June \citep{nakahira2014}.  (See Section 3.1.)  Thus hard tails might be common in NS hard states, but detecting them requires long exposure times with the current instrument sensitivities.

With detections of high-energy tails in NS hard states, their hard X-ray/soft gamma-ray spectral shapes in the hard state show another similarity to BH spectra, as hard tails have been reported for a number of sources (e.g. Cyg X-1, GRS 1915+105, GX 339-4).  These commonalities suggest that the same physical mechanisms are at work in both types of sources.  An interesting difference, though, is at radio wavelengths where emission is high compared to the soft state for BH while for NS the emission is low compared to the soft state.

\section*{Acknowledgments}  The \textit{INTEGRAL} SPI project has been completed under the responsibility and leadership of CNES.  We are grateful to ASI, CEA, CNES, DLR, ESA, INTA, NASA and OSTC for support.

\end{document}